\documentclass[aps,prl,preprint,superscriptaddress]{revtex4-1}
\usepackage{times}
\usepackage{graphicx}
\usepackage[version=3]{mhchem} 
\usepackage[T1]{fontenc}       
\usepackage[colorlinks=true,linkcolor=blue,citecolor=blue]{hyperref}
\usepackage{soul}

\def\unit#1{\mathord{\thinspace\rm #1}}
\def\func#1{\mathop{\rm #1}\nolimits}

\begin{document}
\title{Guiding of Electrons in a Few Mode Ballistic Graphene Channel}
\author{Peter Rickhaus}
\affiliation{Department of Physics, University of Basel, Klingelbergstrasse 82, CH-4056 Basel, Switzerland}

\author{Ming-Hao Liu}
\affiliation{Institut f\"ur Theoretische Physik, Universit\"at Regensburg, D-93040 Regensburg, Germany}

\author{P\'eter Makk}
\email{peter.rickhaus@unibas.ch}
\affiliation{Department of Physics, University of Basel, Klingelbergstrasse 82, CH-4056 Basel, Switzerland}

\author{Romain Maurand}
\affiliation{Department of Physics, University of Basel, Klingelbergstrasse 82, CH-4056 Basel, Switzerland}
\altaffiliation{University Grenoble Alpes, CEA-INAC-SPSMS, F-38000 Grenoble, France}

\author{Samuel Hess}
\affiliation{Department of Physics, University of Basel, Klingelbergstrasse 82, CH-4056 Basel, Switzerland}

\author{Simon Zihlmann}
\affiliation{Department of Physics, University of Basel, Klingelbergstrasse 82, CH-4056 Basel, Switzerland}

\author{Markus Weiss}
\affiliation{Department of Physics, University of Basel, Klingelbergstrasse 82, CH-4056 Basel, Switzerland}

\author{Klaus Richter}
\affiliation{Institut f\"ur Theoretische Physik, Universit\"at Regensburg, D-93040 Regensburg, Germany}

\author{Christian Sch\"onenberger}
\affiliation{Department of Physics, University of Basel, Klingelbergstrasse 82, CH-4056 Basel, Switzerland}

\date{\today}

\begin{abstract}In graphene, the extremely fast charge carriers can be controlled by electron-optical elements, such as waveguides, in which the transmissivity is tuned by the wavelength. In this work charge carriers are guided in a suspended ballistic few-mode graphene channel, defined by electrostatic gating. By depleting the channel, a reduction of mode number and steps in the conductance are observed, until the channel is completely emptied. The measurements are supported by tight-binding transport calculations including the full electrostatics of the sample.\newline\\
\href{http://dx.doi.org/10.1021/acs.nanolett.5b01877}{DOI:10.1021/acs.nanolett.5b01877}

\end{abstract}
\maketitle

%
%
%


The wavelength of electrons at the Fermi energy in graphene \cite{Novoselov2005} can be tuned locally by changing the carrier density $n$ with gate-electrodes \cite{Huard2007,Gorbachev2008}. By forming regions of different wavelengths $\lambda=\sqrt{4\pi/|n|}$, elements known from optics such as lenses \cite{Cheianov2007}, filters \cite{Katsnelson2006} or Fabry-P\'erot interferometers \cite{Miao2007,Rickhaus2013,Grushina2013} can be mimicked. The reflection, refraction and transmission behavior of graphene's massless charge carriers in a spatially varying potential is analogous to the propagation of an electromagnetic wave in media with varying refractive index. An assembly of such media is used to guide photons in optical fibers (OFs). These consist of materials that are assembled in a way that the refractive index in the light-carrying core is larger than the refractive index in the coating. Similar conditions can be achieved in graphene using local gates. The possibility to form electronic waveguides in graphene has attracted a lot of theoretical interest.\cite{Pereira2006,Beenakker2008,Hanson2008,Zhang2009,Villegas2010,Hartmann2010,Wu2011,Stone2012}
However, up to now experiments were performed only in diffusive samples \cite{Williams2011}, making the comparison to optics difficult. In addition, a graphene waveguide is expected to show a distinct behavior once a p-n interface is involved. At a p-n interface the polarity of charge carriers is inverting. Due to the additional confinement of such an interface it is possible to transport one or few modes in a waveguide. In optics, single mode fibers are used for long distant communication since they are not limited by modal dispersion. In a similar way single or few-mode electron-optic fibers can be beneficial for quantum information communication. 

So far, confinement of charge carriers has been mainly achieved with hard-wall potentials, as provided by the edges in graphene-nanoribbons\cite{Wang2008,Tombros22011,Baringhaus2014} or the induced gap in bilayer graphene\cite{Goossens2012,Allen2012}. Whereas the former suffers from irregular edges, the latter is performed in gapped bilayer graphene which does not host relativistic Dirac particles. Even though electrons were successfully guided in the bulk of single-layer graphene by making use of snake states\cite{Williams22011,Rickhaus2015,Taychatanapat2015} their transport properties is significantly altered by the involved perpendicular magnetic field. It is a different and very challenging task to confine the charge carriers in an electrostatic waveguide \cite{Hartmann2010,Stone2012}. In this letter we report on the formation of a narrow tunable ballistic electrostatic channel in graphene, that can be operated as an optical fiber and features controllable single mode filling.

\begin{figure}[htbp]
    \centering
      \includegraphics[width=1\textwidth]{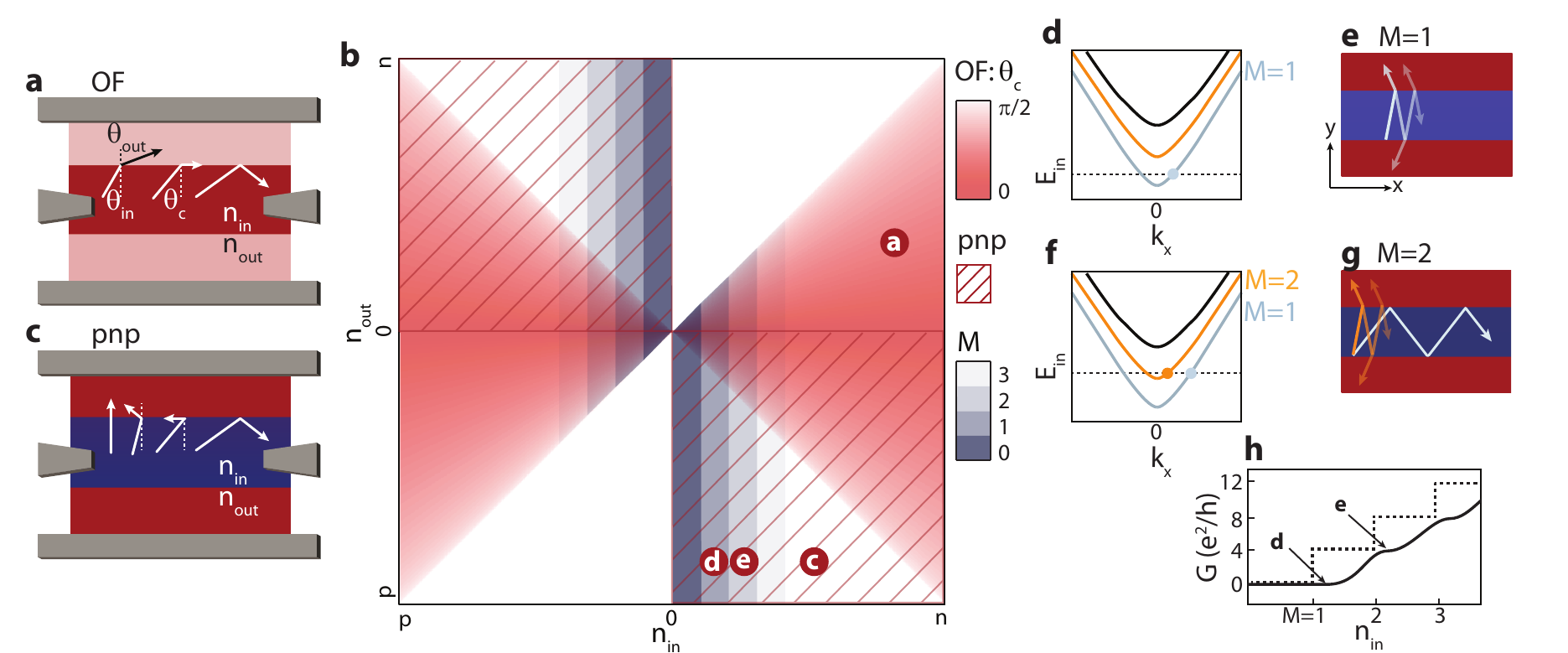}
    \caption{
	\textbf{Guiding mechanisms related to densities in $n_{\rm in}$ and outside $n_{\rm out}$ of an electrostatic channel.}
          \textbf{a}, Possible design of a guiding channel where the gray areas are electrical contacts. If the electron density inside a channel is larger than outside $n_{\rm in}>n_{\rm out}$, electron trajectories reaching the interface under an angle $\theta_{\rm in}$ larger than the angle of total internal reflection $\theta_{\rm c}$ stay in the channel, similar to an optical fiber (OF). 
	  \textbf{b}, The different guiding mechanisms are sketched depending on $n_{\rm in}$ and $n_{\rm out}$. In the red shaded triangles OF guiding is present. The shading visualized $\theta_{\rm c}(n_{\rm in},n_{\rm out})$. The hatched quadrants indicate the regions of pnp-guiding and the blue steps correspond to different mode number $M$ in the channel.
      \textbf{c}, The formation of a p-n junction helps to keep charge carriers in the channel; loss is due to trajectories with (almost) zero angle of incidence.
     \textbf{d}, Sketch of a band diagram in the channel for the situation, when the first mode ($M=1$) is populated. The small value for $k_{\rm x}$ will lead to leakage as sketched in \textbf{e}.
     \textbf{f}, By increasing the Fermi-energy $E_{\rm in}$ the second mode becomes available and $k_{\rm x}$ of the first mode is increased. The first mode can be guided in the channel, as illustrated in \textbf{g}.
     \textbf{h}, Due to this mechanism, the expected conductance plateaus will be smoothened and shifted with respect to $M$ compared to the situation of a hard-wall channel (dashed line). They are still expected to occur at values of $4\:e^2/h$.
}
    \label{guid:figure1}
\end{figure}

In single layer graphene, the local Fermi energy $E=\hbar v_{\rm F}\sqrt{|n|\pi}$, with $v_{\rm F}\approx 10^{6}\unit{m s^{-1}}$ the Fermi velocity in graphene, takes the role of the refractive index. By changing this energy using local gates, a waveguide can be formed. This guiding principle and a possible geometry is shown in Figure \ref{guid:figure1}a. The refraction at an interface is described by Snell's law in graphene: $E_{\rm in}\cdot\sin(\theta_{\rm in})=E_{\rm out}\cdot\sin(\theta_{\rm out})$ and the critical angle for total internal reflection is simply given by $\theta_{\rm c}=\arcsin(E_{\rm out}/E_{\rm in})=\arcsin(\sqrt{|n_{\rm out}|/|n_{\rm in}|})$. For small density ratios $|n_{\rm out}|/|n_{\rm in}|$, the resulting small $\theta_{\rm c}$ will keep electron trajectories very efficiently in the channel. In Figure \ref{guid:figure1}b we plot $\theta_{\rm c}$ as a function of the densities in- and outside of a channel. Negative densities correspond to hole-like, positive to electron-like transport. The resulting red shaded triangles correspond to the region where OF guiding is possible. 

In addition to this OF guiding, charge carriers in graphene can also propagate in a channel that has interfaces at which the polarity of the charge carrier is inverted. Such a channel can be formed by tuning the channel region to electron-like (n) and the outer region to hole-like (p) doping, or \textit{vice versa} as indicated by the hatched quadrants in Figure \ref{guid:figure1}b \cite{Williams2011}. Since at a p-n interface the density is zero, it is naturally reflective. If the transition from p- to n-doping is gradual, the low-density region is increased in size leading to an even more reflective interface. In other words, a smooth p-n interface can guide electrons more efficiently than a sharp one. Losses out of a p-n channel are in both cases mostly caused by trajectories perpendicular to the interface that are transmitted with probability one (Klein tunneling) \cite{Young2009,Cheianov2006,Katsnelson2006} as sketched in Figure \ref{guid:figure1}c. 

At low densities $n_{\rm in}$, the wavelength in the channel is such that $\lambda/2$ becomes larger than the channel width $W$. This situation of a depleted channel is sketched in the dark blue region in Figure \ref{guid:figure1}b. By increasing $n_{\rm in}$, the local Fermi-energy $E_{\rm in}$ increases such that the first mode $M=1$ in the channel can be populated, as shown in the sketch of the band-diagram in Figure \ref{guid:figure1}d. Yet, due to the low value of the wavevector $k_{\rm x}$ in propagation direction $x$, the angle of incidence at the p-n interface is close to perpendicular and the mode will leak out of the channel (Figure \ref{guid:figure1}e) and cannot be observed in transport along the channel. Nevertheless, by increasing $E_{\rm in}$, $k_{\rm x}$ increases and the angle of incidence towards the p-n interface becomes larger. A corresponding band-diagram is sketched in Figure \ref{guid:figure1}f, where already the second mode ($M=2$) is available. In this situation, the first mode can be guided and the second mode leaks out of the channel since the Fermi-energy crosses this band at low $k_{\rm x}$. Compared to a channel where the outer regions are forbidden for electrons, the expected conductance plateaus for transport along the channel will be smoothened due to the above discussed mechanism. We still expect conductance steps at $4e^2/h$, as sketched in Figure \ref{guid:figure1}h. A more detailed analysis is given in the supplementary material. 



\begin{figure}[htbp]
    \centering
      \includegraphics[width=1\textwidth]{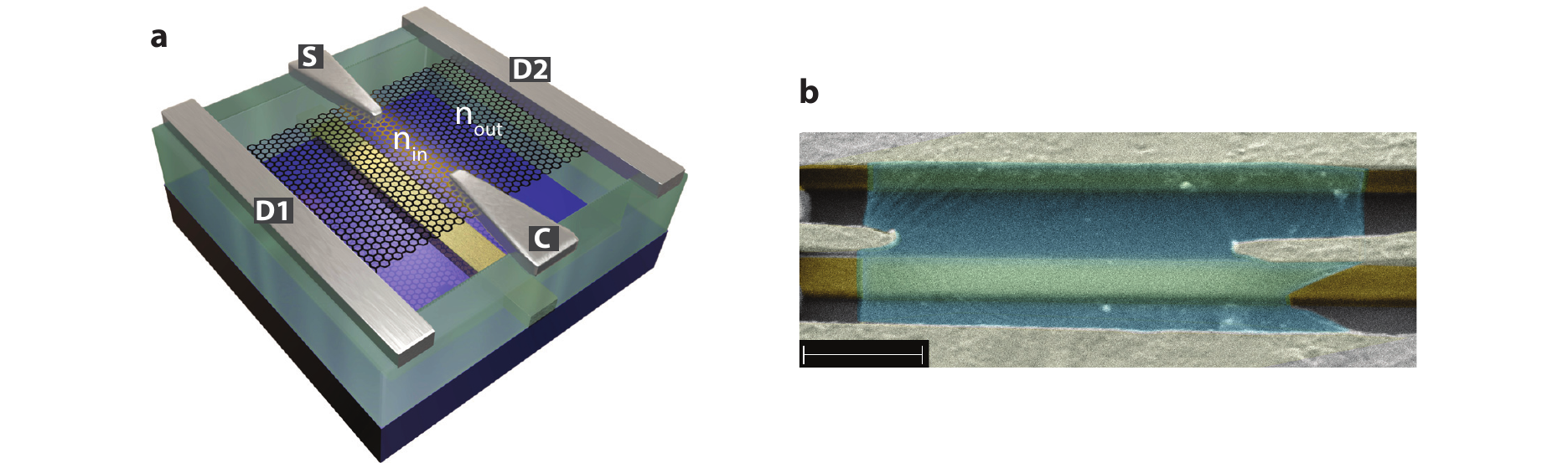}
    \caption{
	\textbf{Realization of a guiding channel in a suspended graphene device.}
	\textbf{a}, 3-dimensional design of the device. Tuning the density in the channel ($n_{\rm in}$) with the bottomgate (gold) and the density outside of the channel ($n_{\rm out}$) with the backgate (blue), the electrons can be guided between the source (S) and the collector (C) contacts. The losses are recorded at side drain contacts D1 and D2.
	\textbf{b}, False-colored SEM image of the measured device; scale-bar: $1\:\mu$m. 
}
    \label{guid:figure2}
\end{figure}

We realized an electrostatic channel in a four-terminal suspended graphene device. A 3-dimensional design is shown in Figure \ref{guid:figure2}a. By applying a voltage on the backgate (blue) and bottomgate (gold), the density can be tuned locally to guide electrons between the source (S) and collector (C) contact. The guiding losses can be measured at the side drain contacts (D1, D2). A scanning electron micrograph of the measured device is given in Figure \ref{guid:figure2}b. The graphene flake is roughly $4\times2\:{\rm\mu m}^2$ in size. The fabrication follows the recipe of Refs. \citenum{Tombros2011,Maurand2014}. Graphene is transferred onto a sacrificial polymer layer (LOR) using a dry-transfer technique, and after defining the contacts, suspended. It is cleaned by in-situ current annealing between the D1 and D2 contact.

\begin{figure}[htbp]
    \centering
      \includegraphics[width=1\textwidth]{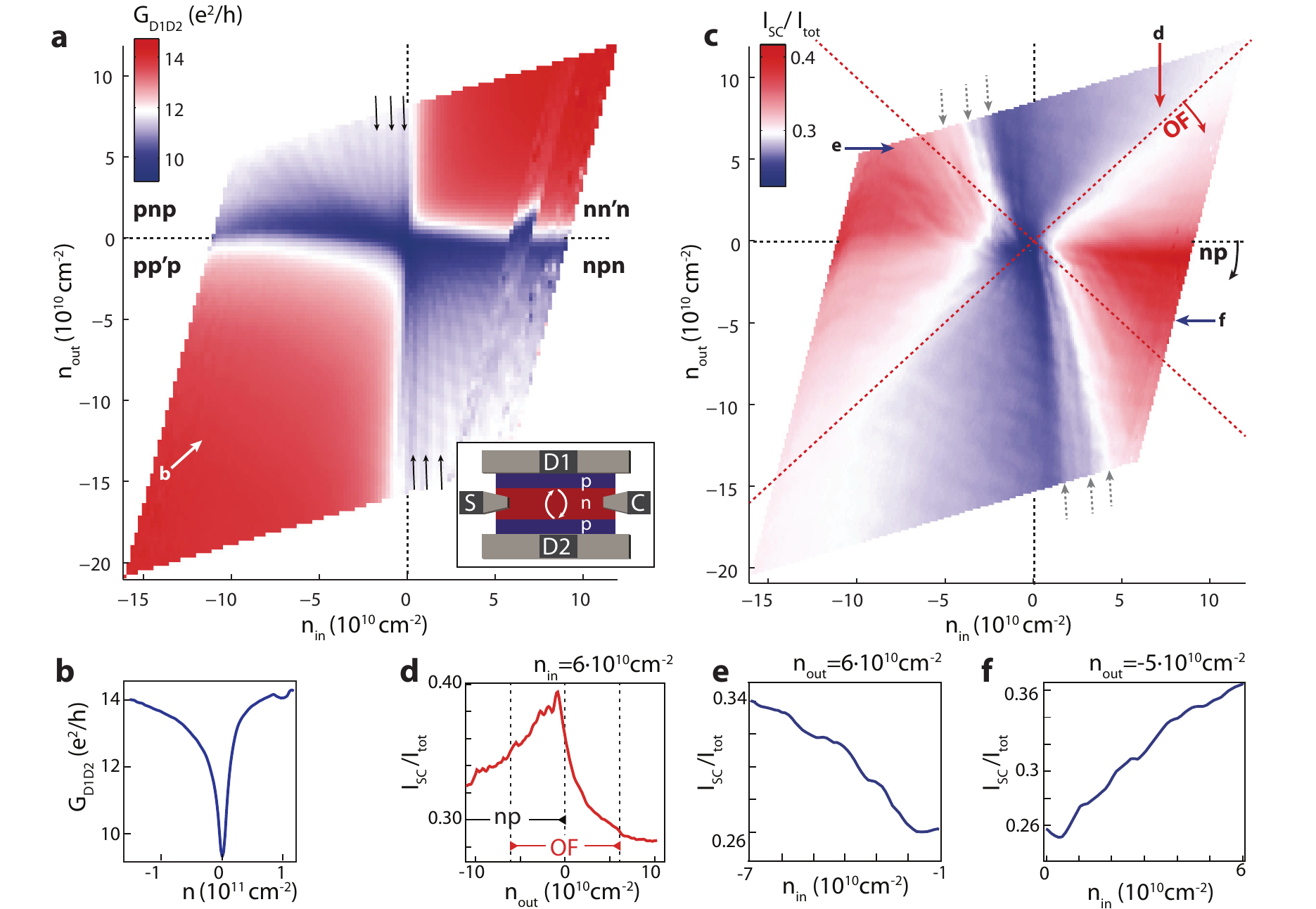}
    \caption{
	\textbf{Experimental results: Fabry-P\'erot interferences across the channel and guiding efficiency along the channel.}
      \textbf{a}, Two-terminal conductance $G_{\rm D1D2}$ between D1 and D2 as a function of the densities $n_{\rm in}$ and $n_{\rm out}$ (S and C are floating). The resonances marked with black arrows are due to Fabry-P\'erot interferences in the channel as sketched in the inset. 
      \textbf{b}, Field-effect as a function of uniform doping (cut along white arrow in a).
      \textbf{c}, Guiding efficiency $I_{\rm SC}/I_{\rm tot}$ as a function of $n_{\rm in}$ and $n_{\rm out}$. Here, an AC voltage is applied to the S contact and current is measured simultaneously at C and D1, D2.
      \textbf{d}, Cut in c at fixed $n_{\rm in}=6\cdot10^{10}\rm{cm}^{-2}$. The guiding efficiency increases drastically in the OF regime and is highest when pn and OF guiding coexist.
      \textbf{e}, The cut in c around $n_{\rm out}=6\cdot10^{10}\rm{cm}^{-2}$ reveals steps in the guiding efficiency that are due to mode filling of the channel. Shown is an averaged curve of $n_{\rm out}=5...7\cdot10^{10}\rm{cm}^{-2}$. The steps are seen in the 2D plot in c and are marked there with gray vertical arrows. 
	\textbf{f}, A similar cut at  $n_{\rm out}=-4...-6\cdot10^{10}\rm{cm}^{-2}$ reveals steps also for inverted polarity.
		}
    \label{guid:figure3}
\end{figure}


As a first characterization the conductance $G_{\rm D1D2}$ between D1 and D2 is measured while the S and C contacts are floating. We correct for the capacitive cross talk between backgate and bottomgate voltage and plot the data as a function of channel densities $n_{\rm in}$ and $n_{\rm out}$. The conversion from voltage to density is given by the gating efficiency which we extracted from the electrostatic simulation (for details, see supplementary material). Throughout this paper negative densities correspond to holes, positive to electron charge carriers. In the colorscale-map of Figure \ref{guid:figure3}a, the conductance drops drastically once a p-n junction is formed (i.e. the blue regions). Figure \ref{guid:figure3}b shows the field effect along uniform doping which reveals a pronounced Dirac point. The very sharp transition from the unipolar to the bipolar region occurs within $n_{\rm in}\approx 10^9\;\rm{cm}^{-2}$. This demonstrates the very high quality of the graphene. In addition, the regular Fabry-P\'erot pattern (marked with arrows and sketched in the inset of Figure \ref{guid:figure3}a) indicates ballistic transport in the channel \cite{Rickhaus2013,Grushina2013}.

We now discuss the guiding efficiency measurement (fig. \ref{guid:figure3}c). An AC voltage is applied to the S contact and current is measured at C, D1 and D2 using an IV-converter and a Lock-In detector on each terminal. The guiding efficiency $\gamma=I_{\rm SC}/I_{\rm tot}$ is then given by the current $I_{\rm SC}$ measured at C  divided by the total current $I_{\rm tot}=I_{\rm SD1}+I_{\rm SD2}+I_{\rm SC}$. It is important to notice that $\gamma$ includes the injection efficiency into the channel which is limiting the overall maximum efficiency to $42\%$ in this device. In this complex geometry it is not possible to discriminate the losses of the channel itself from losses due to poor coupling of the charge carriers into the channel. The colorscale map $\gamma(n_{\rm in},n_{\rm out})$ shows high guiding efficiency in the expected regions: A large $\gamma$ is observed in the OF triangles, and the efficiency is even increased once a p-n-junction is formed. This is best seen by taking a cut at $n_{\rm in}=6\cdot10^{10} \:\rm{cm}^{-2}$ shown in Figure \ref{guid:figure3}d and indicated by a red arrow in Figure \ref{guid:figure3}c. Charge carriers cannot be guided for $n_{\rm out}>n_{\rm in}$ and for uniform doping ($n_{\rm out}=n_{\rm in}$) roughly 29\% of the charge carriers reach the C contact. In the OF regime, the efficiency increases drastically and reaches its maximum if the outer region is depleted, i.e. $n_{\rm out}\approx 0$. The formation of a p-n-interface leads to an increased $\gamma$ compared to the regime where only OF guiding is present. Even though OF and p-n guiding mechanisms have been discussed for a diffusive sample in Ref. \cite{Williams2011} here we demonstrate their occurrence in the expected regions in a density-density map of a ballistic sample. 

The lowest efficiency in the colorscale map (fig. \ref{guid:figure3}c) is observed in the empty-channel region (dark blue). A step-wise increase of $\gamma$ starting from the depleted channel towards larger $n_{\rm in}$ is seen in the colorscale map. By taking cuts around $n_{\rm out}=6\cdot10^{10}\:\rm{cm}^{-2}$  plateaus corresponding to a change in mode number $M$ become apparent (Figure \ref{guid:figure3}e). For this cut we averaged the curves between $n_{\rm out}=5...7\cdot10^{10}\:\rm{cm}^{-2}$ to wash out features that are changing with $n_{\rm out}$. More details on these features are given later in the manuscript. The plateaus are also present at the opposite polarity, i.e. an n-doped channel and a p-doped outer region. They can be seen for example in cuts around $n_{\rm out}=-5\cdot10^{10}\:\rm{cm}^{-2}$ as given in Figure \ref{guid:figure3}f.

\begin{figure}[htbp]
    \centering
      \includegraphics[width=1\textwidth]{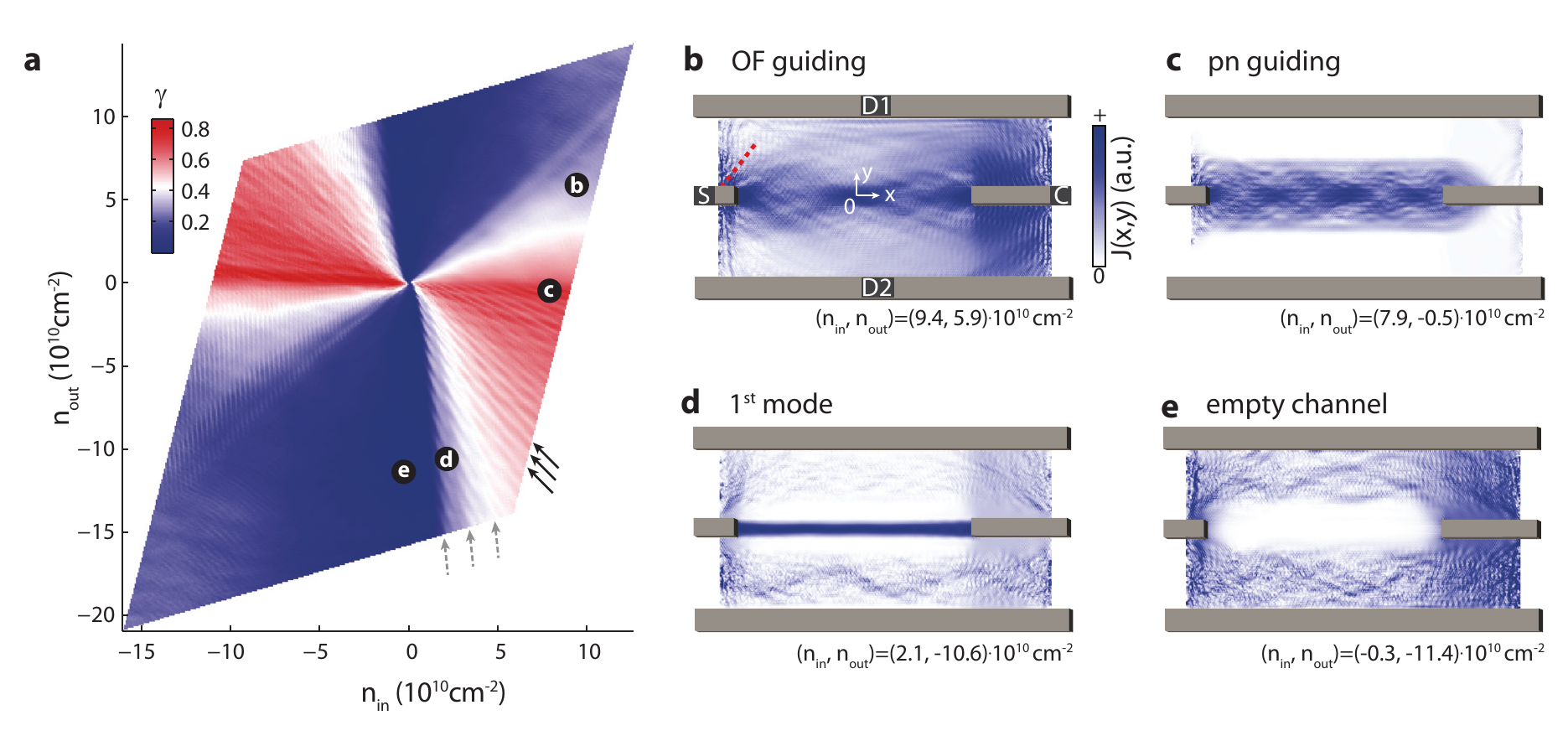}
    \caption{
	\textbf{Simulated guiding efficiency and local current density profiles based on tight-binding model.}
      \textbf{a}, Calculated guiding efficiency following the same definition as the experiment, $\gamma= I_{\rm SC}/I_{\rm tot}$, using 3-dimensional electrostatics given by the geometry of the device. Sequential mode filling of the channel is indicated by gray dashed lines. Black arrows mark a resonance pattern involving the inner cavity formed by the narrow channel and an outer cavity given by one outer contact D1,D2 and the channel. 
	\textbf{b}, Local current density distribution for a small applied voltage on the source contact S. The densities are tuned to the optical fiber regime. The red dashed line shows the angle of total internal reflection. 
	\textbf{c}, In the p-n-regime, the electrons are kept inside the channel very efficiently and complex resonance patterns are observed. Close to the S contact current is lost towards D1 and D2 due to Klein tunneling. 
	\textbf{d}, Local current distribution for the first mode.
	\textbf{e}, At vanishing $n_{\rm in}$, the channel can be emptied, i.e., the local current density is zero inside the channel.
		}
   \label{guid:figure4}
\end{figure}

For a comprehensive understanding, we compare the experiment to transport calculations based on an ideal electrostatic model obtained by finite-element simulations following the gating and graphene flake geometry of our device. Using the scalable tight-binding model for graphene \cite{Liu2015}, the full $4\times2\;{\rm\mu m}^2$ flake with realistic on-site energy profiles $V(x,y)$ from the electrostatics can be considered and the main features of the guiding efficiency map can be reproduced by applying the real-space Green's function method to compute the conductance. The results of the simulations are shown in Figure \ref{guid:figure4}a. The main features of the measurement are well captured: There is a high guiding efficiency in the OF triangles which is even increased in the p-n-p-regime. We see a very low efficiency at low $n_{\rm in}$ where the channel is empty.

The emerging modes confined in the channel are observed along lines indicated by gray dashed arrows in Figure \ref{guid:figure4}a. Additional resonances (indicated by black arrows) in the p-n-p regions appear and will be referred to as two-cavity resonances. They are related to states spatially extending over both the inner channel and the cavity regions confined by D1 and D2. They are therefore influenced by both $n_{\rm in}$ and $n_{\rm out}$ and are roughly parallel to the anti-diagonal in the density map. Further analysis is given in the supplementary material.

From the simulations we further extract the space-resolved local current density $\mathbf{J}=J_x\mathbf{e}_x+J_y\mathbf{e}_y$ in the different guiding situations by applying a small DC voltage difference between the S and C contact with D1 and D2 grounded. In Figure \ref{guid:figure4}b we plot $J(x,y)=[J_x^2(x,y)+J_y^2(x,y)]^{1/2}$ in the optical fiber regime ($|n_{\rm in}|>|n_{\rm out}|$). Most current propagates in a channel towards the right contact. The loss (i.e. current towards D1 or D2) is given by trajectories injected at S that reach the interface close to normal incidence. For the given densities we obtain $\theta_{\rm c}=52^{\circ}$ and we sketch a corresponding dashed line in the Figure.
By the formation of a p-n-interface this loss is reduced drastically as shown in Figure \ref{guid:figure4}c. Outside the channel, the current is almost completely suppressed. Only close to the injector contact a loss current is observed, corresponding to trajectories aligned perpendicular to the channel border. For smooth p-n-junctions, this is exactly what is expected \cite{Cheianov2006} due to strong Klein collimation. Inside the p-n-channel, a complex interference pattern forms. By reducing the density in the channel only the first mode is filled (Figure \ref{guid:figure4}d) and as a consequence the complex interference pattern of Figure \ref{guid:figure4}c disappears. Even though the channel appears to be lossless, the injection into the single-mode channel is rather inefficient, i.e. there is a number of modes outside the channel that can be populated. For this reason the current density in the outer cavity does not disappear completely. By further reducing $n_{\rm in}$, the channel can be emptied completely, as it becomes apparent in Figure \ref{guid:figure4}e.

\begin{figure}[htbp]
    \centering
      \includegraphics[width=1\textwidth]{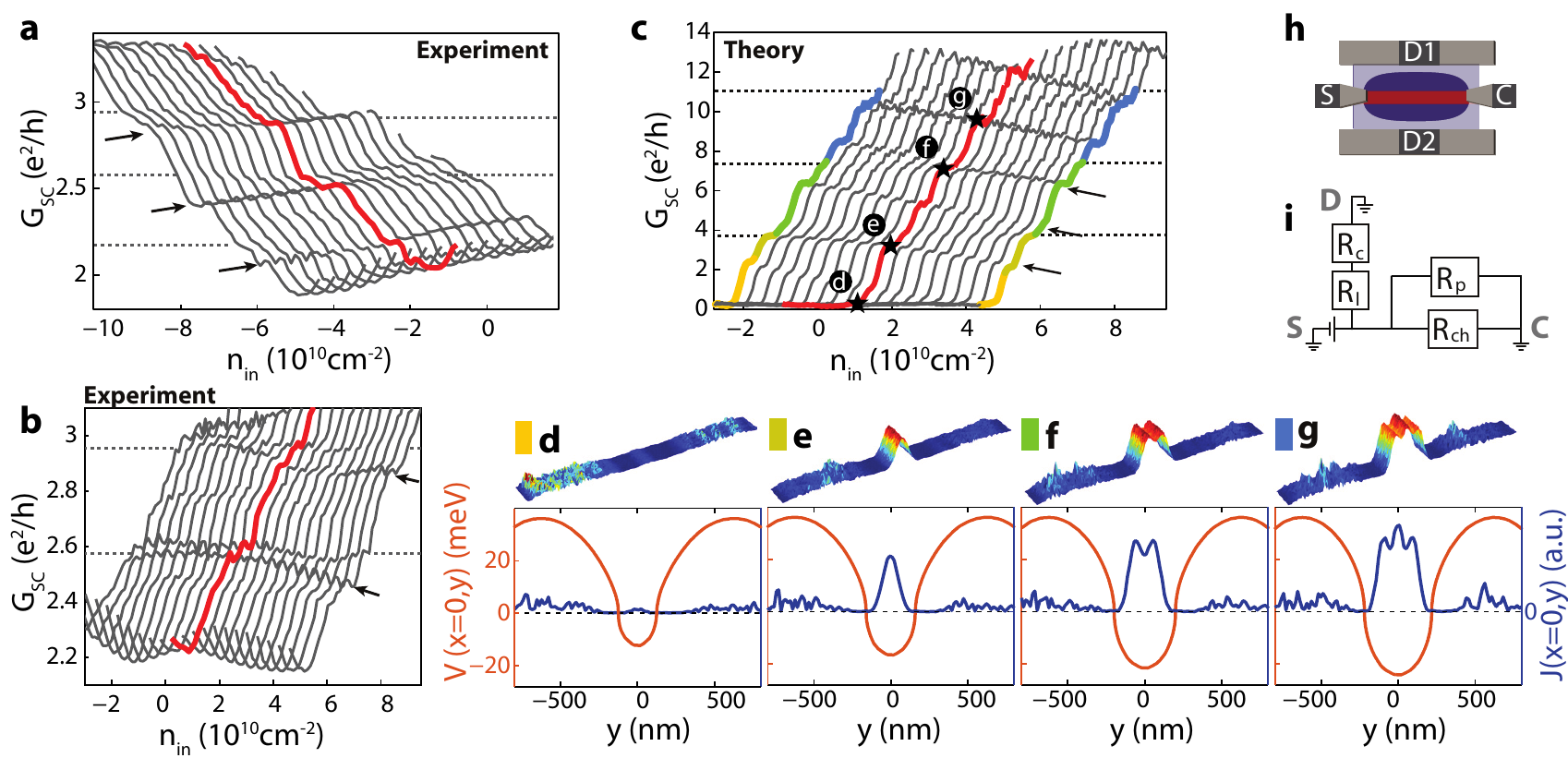}
    \caption{
	\textbf{Discrete change of the number of modes in experimental and theoretical conductance data.}
     \textbf{a}, Steps corresponding to mode-filling are seen in the experimental $G_{\rm SC}(n_{\rm in})$ data at fixed $n_{\rm out}=5...7\cdot 10^{10}\rm{cm}^{-2}$. The curves are offset from the red line by $n_{\rm in}=0.4\cdot 10^{10}\rm{cm}^{-2}$. Mode filling (see Figure \ref{guid:figure1}h) occurs along the dashed lines and an additional two-cavity resonance pattern is indicated by the black arrows.
      \textbf{b}, A comparable pattern, yet less clear, is observed for inverted polarity.
      \textbf{c}, The calculated $G_{\rm SC}(n_{\rm in})$ curves reveal the plateaus, indicated by black dashed lines, and the two-cavity resonance pattern (black arrows).  
        \textbf{d}, Local band offset $V(y)$ (orange curve) at the middle of the channel $x=0$. Blue curve: corresponding local current density $J(x=0,y)$. Top panel: $J(x,y)$ for $x$ close to $0$. For the given density profile the channel could carry one single mode $M=1$, which however is leaking out drastically such that almost no current is transported. The plots d-g are taken at the density marked by a star in c but are representative for the respective conductance regimes marked by colors in the outer two curves in c.
      \textbf{e}, As the channel gets deeper and wider, the first mode can be transported in the channel.
      \textbf{f},\textbf{g}, The second and third mode appear when $|n_{\rm in}|$ increases further.
 \textbf{h},\textbf{i}, A model explaining the occurrence of the conductance plateaus at $2.2$, $2.6$ and $2.9e^2/h$ in \textbf{a}. Regions that are doped by absorbates are colored in light blue. The channel resistance $R_{\rm ch}$ is then not only given by $1/(N\cdot4e^2/h)$ but modified by a transmission probability $t$ into and out of the channel. Additionally, current can flow from S to C through the outer cavities, which is modeled by a parallel resistor $R_{\rm p}$. The measured resistance between S and C is then given by $R=(1/(t^2\cdot N\cdot4e^2/h)+R_{\rm p})^{-1}$. The loss resistance from S to D, $R_{\rm l}$, is modified by a contact resistance in series $R_{\rm c}$. If we calculate $G'_{\rm SC}=1/R_{\rm ch}$ at $n_{\rm out}=5\cdot 10^{10}\rm{cm}^{-2}$ for $R_{\rm p}=13.4k\Omega$ and $t=0.32$ the plateaus occur at $0,\:4,\:8$ and $12e^2/h$.
      }
    \label{guid:figure5}
\end{figure}


We now show in figure \ref{guid:figure5}a-c the conductance between source and collector in the interesting region of biploar doping where the channel is close to full depletion and plateaus in the conductance can be expected due to quantized filling. In Figure \ref{guid:figure5}a and b several curves $G_{\rm SC}(n_{\rm in})$ are given for values of $n_{\rm out}=5...7\cdot10^{10}$ cm$^{-2}$. They are offset from the red curve by multiples of $n_{\rm in}=0.4\cdot10^{10}$ cm$^{-2}$. The plateaus due to mode-filling are indicated by dashed lines and we mark the additional two-cavity resonances by the black arrows. A similar structure is observed for opposite polarity (Figure \ref{guid:figure5}b). 

Although Fabry-P\'erot resonances and mode filling occur on similar length and energy scales they can be distinguished by their different gate behaviour.
The observed pattern is reproduced clearly in the tight binding simulation for $G_{\rm SC}(n_{\rm in})$ shown in Figure \ref{guid:figure5}c. Again the flat plateaus are marked with dashed lines and the dispersing resonances are indicated by arrows. Comparing to Figure \ref{guid:figure4}a we confirm that the resonances indicated by the black arrows correspond to the above mentioned two-cavity resonances whereas the plateaus indicated by horizontal dashed lines are due to an increase of the channel mode-number. These plateaus occur at roughly 4, 8 and 12 $e^2/h$ here. Compared to the two-cavity resonances they do not disappear if the side contacts (D1 and D2) are reflective, as we reveal in the supporting material, where we also show steps in the conductance for a second device. 
The number of modes $M$ in the middle of the channel $(x=0)$ can be read off from the number of peaks in $J(x=0,y)$ in Figures \ref{guid:figure5}d-g, where the potential landscape $V(x=0,y)$ from the electrostatics is also shown. On top the $J(x,y)$ profiles for $x$ close to $0$ are given. The positions where the profiles were imaged are labeled with stars in Figure \ref{guid:figure5}c, and their shape is representative for the whole respective conductance range marked by colored lines in Figure \ref{guid:figure5}c. The $V(x=0,y)$ profiles reveal that the density is gradually changing from n outside to p inside of the channel. The smoothness in the potential landscape is due to the large distance of $\approx400$ nm between the graphene and gate electrodes.
The channel formed by the potential in Figure \ref{guid:figure5}d can transport only one mode, but since $k_x$ is small and since the channel is very leaky for small angles as explained in Figure \ref{guid:figure1}d-e., only a vanishing current is observed in the middle. By increasing the voltage on the inner gate, the minimum in the potential profile $V(x=0,y)$ decreases and the channel grows wider such that electrons gain momentum in $x$-direction and the first mode can be transported along the waveguide (Figure \ref{guid:figure5}e). By further increasing the inner gate voltage, the second mode, $M=2$ (Figure \ref{guid:figure5}f), and third mode, $M=3$ (Figure \ref{guid:figure5}g), also become available for charge transport along the channel.
  
Having confirmed the origin of the non-dispersing features in the experiment we now compare the conductance values. In the experiment, the plateaus do not occur at $G'_{\rm SC}=N\cdot4e^2/h$ (with $N=0,1,2,...$) but rather at $G_{\rm SC}=1.9,\:2.2,\:2.6$ and $2.9e^2/h$ in Figure \ref{guid:figure5}a. A simple model, depicted in Figure \ref{guid:figure5}h-i, explains this deviation by taking into account non-uniform cleanliness of the graphene. In Figure \ref{guid:figure5}h regions that presumably are doped by absorbents on the surface are shaded with light-blue color. Such a distribution of dopants is likely to occur after current annealing since the contacts act as heat sinks and therefore attract residual absorbents when the graphene is hot \cite{Freitag2012}. 
In a simple resistor network (Figure \ref{guid:figure5}i) we therefore model an additional resistor $R_{\rm p}$, due to trajectoires from S to C that do not penetrate the channel, in parallel to the channel resistance $R_{\rm ch}$. In addition the injection into the channel and the detection at the collector will be modified by a transmission probability $t$ which is the transmission from the contact into and out of the channel. Therefore, $R_{\rm ch}=1/(t^2\cdot N\cdot4e^2/h)$ and the measured resistance between S and C will be $R=(1/R_{\rm ch}+1/R_{\rm p})^{-1}=1/G_{\rm SC}$. For the $G_{\rm SC}(n_{\rm in})$ curve at $n_{\rm out}=5\cdot 10^{10}\rm{cm}^{-2}$ plateaus at $G'_{\rm SC}=0,\:4,\:8$ and $12e^2/h$ are obtained for $R_{\rm p}=13.4k\Omega$ and $t=0.32$. 
By subtracting a contact resistance $R_{\rm c}=2.9k\Omega$ from the loss conductance $G_{\rm SD1D2}$, we find guiding efficiencies $\gamma=G'_{\rm SC}/(G'_{\rm SC}+G'_{\rm SD1D2})$ of $26\%,41\%$ and $51\%$ for the first, second and third plateau. $R_{\rm c}$ is extracted for high, unipolar densities in the $G_{\rm SD1D2}$ map (shown in the supporting material). The values compare well to the simulation (Figure \ref{guid:figure4}a) where  $28\%,39\%$ and $46\%$  are obtained.

Apparently the main limitation in the experiment is the injection and detection efficiency, which is parametrized in $t$. It is with $t=0.32$ a factor of $3$ smaller than in the simulation, where $t=0.96$ is found by applying the same resistor model.
Improvements would focus on creating excellent contact to the graphene. This would, on one hand, increase $t$ both at the source and collector contact. On the other hand, a reduction of $R_{\rm p}$ could be expected since the D1, D2 contacts would be less reflective, i.e. trajectories from S that get reflected at D1 and enter C could be suppressed. A strategy to increase $t$ by shaping the graphene is discussed in the supplementary material. 

In the simulation, however, $t$ is large meaning that the ideal channel very efficiently transports modes once $k_x$ is sufficiently large. This efficient guiding is only possible due to the smooth p-n interface in our device \cite{Hartmann2010} which is very reflective for larger angles of incidence $\theta$. For comparison, a sharp p-n interface transmits trajectories with $\theta=45^\circ$ with $50\%$ probability \cite{Cheianov2006} such that the mode sketched in Figure \ref{guid:figure1}g could not be guided. For our smooth interface, on the other hand, $50\%$ transmission is obtained for only $\theta\approx20^\circ$, and at $45^\circ$ the transmission is close to zero. These values are taken for $(n_{\rm in},n_{\rm out})=(5,-10)\cdot10^{10}\;cm^{-2}$. Details are given in the supplementary material.


In conclusion, we have realized electrostatically defined electron waveguides in ballistic graphene with an aspect ratio >10 using a global and a channel gate. Typical channel parameters are $\approx300$ nm in width and $2.7\:\mu$m in length. The wavelengths of the charge carriers in- and outside of the channel can be tuned by the gates independently. Using this tunability, we can distinguish and control the regions of optical fiber- and p-n guiding. We observe an increased guiding efficiency if the two mechanisms coexist. Using p-n guiding, clear steps in the channel conductance appear whenever the mode number changes by one. All experimental results are supported by self-consistent theoretical simulations which also reveal that a smooth p-n interface and a sufficiently large momentum $k_{\rm x}$ along the channel is required to transport single modes. The simulation provides also a powerful guide for future device improvements.  It shows, for example, that by using an even smoother p-n interface to confine the electrons in the channel, modes could be guided for $k_{\rm x}\ll k_{\rm y}$. In current devices the coupling from the injector contact to the channel and similarly from the channel to the detector contact is far from ideal. The injector and collector efficiencies of currently $\approx35\%$  can be increased either by using a p-n interface as a collimator or by etching the graphene flake. With such improvements guiding efficiencies $> 80\%$, as demonstrated by the theory, are in reach. This then allows to explore one-dimensional transport in graphene without the need to etch nanoribbons. The difficulty of the required precise control over the sample edges at the atomic level in nanoribbons can then be circumvented. An electrostatically confined one-dimensional channel within the bulk of graphene could even be oriented intentionally into any preferred crystallographic direction allowing to study the symmetry of the one-dimensional bands. A high degree of confinement can further be used for fast switches with a potential visibility in excess of $80\%$.



\vfill
\textbf{Acknowledgments}
M.-H.L. and K.R. acknowledge financial support by the Deutsche Forschungsgemeinschaft (SFB 689). This work was further funded by the Swiss National Science Foundation, the Swiss Nanoscience Institute, the Swiss NCCR QSIT, the ESF programme Eurographene, the EU FP7 project SE2ND, the ERC Advanced Investigator Grant QUEST, and the EU flagship project graphene.

The authors thank Andreas Baumgartner and Szabolcs Csonka for fruitful discussions.

\textbf{Author contributions}
P.R. and S.H. and S.Z. fabricated the devices. Measurements were performed by P.R., S.H., S.Z., P.M., R.M. and M.W. Simulations were performed by M.-H.L. C.S. and K.R. guided the work. All authors worked on the manuscript.


\newpage

\bibliography{guidingLIB}

\providecommand{\latin}[1]{#1}
\providecommand*\mcitethebibliography{\thebibliography}
\csname @ifundefined\endcsname{endmcitethebibliography}
  {\let\endmcitethebibliography\endthebibliography}{}
\begin{mcitethebibliography}{33}
\providecommand*\natexlab[1]{#1}
\providecommand*\mciteSetBstSublistMode[1]{}
\providecommand*\mciteSetBstMaxWidthForm[2]{}
\providecommand*\mciteBstWouldAddEndPuncttrue
  {\def\EndOfBibitem{\unskip.}}
\providecommand*\mciteBstWouldAddEndPunctfalse
  {\let\EndOfBibitem\relax}
\providecommand*\mciteSetBstMidEndSepPunct[3]{}
\providecommand*\mciteSetBstSublistLabelBeginEnd[3]{}
\providecommand*\EndOfBibitem{}
\mciteSetBstSublistMode{f}
\mciteSetBstMaxWidthForm{subitem}{(\alph{mcitesubitemcount})}
\mciteSetBstSublistLabelBeginEnd
  {\mcitemaxwidthsubitemform\space}
  {\relax}
  {\relax}

\bibitem[Novoselov \latin{et~al.}(2005)Novoselov, Geim, Morozov, Jiang,
  Katsnelson, Grigorieva, Dubonos, and Firsov]{Novoselov2005}
Novoselov,~K.~S.; Geim,~A.~K.; Morozov,~S.~V.; Jiang,~D.; Katsnelson,~M.~I.;
  Grigorieva,~I.~V.; Dubonos,~S.~V.; Firsov,~A.~A. \emph{Nature} \textbf{2005},
  \emph{438}, 197--200\relax
\mciteBstWouldAddEndPuncttrue
\mciteSetBstMidEndSepPunct{\mcitedefaultmidpunct}
{\mcitedefaultendpunct}{\mcitedefaultseppunct}\relax
\EndOfBibitem
\bibitem[Huard \latin{et~al.}(2007)Huard, Sulpizio, Stander, Todd, Yang, and
  Goldhaber-Gordon]{Huard2007}
Huard,~B.; Sulpizio,~J.~A.; Stander,~N.; Todd,~K.; Yang,~B.;
  Goldhaber-Gordon,~D. \emph{Physical Review Letters} \textbf{2007}, \emph{98},
  236803\relax
\mciteBstWouldAddEndPuncttrue
\mciteSetBstMidEndSepPunct{\mcitedefaultmidpunct}
{\mcitedefaultendpunct}{\mcitedefaultseppunct}\relax
\EndOfBibitem
\bibitem[Gorbachev \latin{et~al.}(2008)Gorbachev, Mayorov, Savchenko, Horsell,
  and Guinea]{Gorbachev2008}
Gorbachev,~R.~V.; Mayorov,~A.~S.; Savchenko,~A.~K.; Horsell,~D.~W.; Guinea,~F.
  \emph{Nanoletters} \textbf{2008}, \emph{8}, 1995--1999\relax
\mciteBstWouldAddEndPuncttrue
\mciteSetBstMidEndSepPunct{\mcitedefaultmidpunct}
{\mcitedefaultendpunct}{\mcitedefaultseppunct}\relax
\EndOfBibitem
\bibitem[Cheianov \latin{et~al.}(2007)Cheianov, Fal'ko, and
  Altshuler]{Cheianov2007}
Cheianov,~V.~V.; Fal'ko,~V.; Altshuler,~B.~L. \emph{Science} \textbf{2007},
  \emph{315}, 1252--1255\relax
\mciteBstWouldAddEndPuncttrue
\mciteSetBstMidEndSepPunct{\mcitedefaultmidpunct}
{\mcitedefaultendpunct}{\mcitedefaultseppunct}\relax
\EndOfBibitem
\bibitem[Katsnelson \latin{et~al.}(2006)Katsnelson, Novoselov, and
  Geim]{Katsnelson2006}
Katsnelson,~M.~I.; Novoselov,~K.~S.; Geim,~A.~K. \emph{Nature Physics}
  \textbf{2006}, \emph{2}, 620--625\relax
\mciteBstWouldAddEndPuncttrue
\mciteSetBstMidEndSepPunct{\mcitedefaultmidpunct}
{\mcitedefaultendpunct}{\mcitedefaultseppunct}\relax
\EndOfBibitem
\bibitem[Miao \latin{et~al.}(2007)Miao, Wijeratne, Zhang, Coskun, Bao, and
  Lau]{Miao2007}
Miao,~F.; Wijeratne,~S.; Zhang,~Y.; Coskun,~U.~C.; Bao,~W.; Lau,~C.~N.
  \emph{Science} \textbf{2007}, \emph{317}, 1530--1533\relax
\mciteBstWouldAddEndPuncttrue
\mciteSetBstMidEndSepPunct{\mcitedefaultmidpunct}
{\mcitedefaultendpunct}{\mcitedefaultseppunct}\relax
\EndOfBibitem
\bibitem[Rickhaus \latin{et~al.}(2013)Rickhaus, Maurand, Liu, Weiss, Richter,
  and Sch\"{o}nenberger]{Rickhaus2013}
Rickhaus,~P.; Maurand,~R.; Liu,~M.-H.; Weiss,~M.; Richter,~K.;
  Sch\"{o}nenberger,~C. \emph{Nature Communications} \textbf{2013}, \emph{4},
  2342\relax
\mciteBstWouldAddEndPuncttrue
\mciteSetBstMidEndSepPunct{\mcitedefaultmidpunct}
{\mcitedefaultendpunct}{\mcitedefaultseppunct}\relax
\EndOfBibitem
\bibitem[Grushina \latin{et~al.}(2013)Grushina, Ki, and Morpurgo]{Grushina2013}
Grushina,~A.~L.; Ki,~D.-K.; Morpurgo,~A.~F. \emph{Applied Physics Letters}
  \textbf{2013}, \emph{102}, 223102\relax
\mciteBstWouldAddEndPuncttrue
\mciteSetBstMidEndSepPunct{\mcitedefaultmidpunct}
{\mcitedefaultendpunct}{\mcitedefaultseppunct}\relax
\EndOfBibitem
\bibitem[Pereira \latin{et~al.}(2006)Pereira, Mlinar, Peeters, and
  Vasilopoulos]{Pereira2006}
Pereira,~J.~M.; Mlinar,~V.; Peeters,~F.~M.; Vasilopoulos,~P. \emph{Physical
  Review B} \textbf{2006}, \emph{74}, 045424\relax
\mciteBstWouldAddEndPuncttrue
\mciteSetBstMidEndSepPunct{\mcitedefaultmidpunct}
{\mcitedefaultendpunct}{\mcitedefaultseppunct}\relax
\EndOfBibitem
\bibitem[Beenakker \latin{et~al.}(2008)Beenakker, Sepkhanov, Akhmerov, and
  Tworzydlo]{Beenakker2008}
Beenakker,~C. W.~J.; Sepkhanov,~R.~A.; Akhmerov,~A.~R.; Tworzydlo,~J.
  \emph{Physical Review Letters} \textbf{2008}, \emph{102}, 146804\relax
\mciteBstWouldAddEndPuncttrue
\mciteSetBstMidEndSepPunct{\mcitedefaultmidpunct}
{\mcitedefaultendpunct}{\mcitedefaultseppunct}\relax
\EndOfBibitem
\bibitem[Hanson(2008)]{Hanson2008}
Hanson,~G.~W. \emph{Journal of Applied Physics} \textbf{2008}, \emph{103},
  064302\relax
\mciteBstWouldAddEndPuncttrue
\mciteSetBstMidEndSepPunct{\mcitedefaultmidpunct}
{\mcitedefaultendpunct}{\mcitedefaultseppunct}\relax
\EndOfBibitem
\bibitem[Zhang \latin{et~al.}(2009)Zhang, He, and Chen]{Zhang2009}
Zhang,~F.-M.; He,~Y.; Chen,~X. \emph{Applied Physics Letters} \textbf{2009},
  \emph{94}, 212105\relax
\mciteBstWouldAddEndPuncttrue
\mciteSetBstMidEndSepPunct{\mcitedefaultmidpunct}
{\mcitedefaultendpunct}{\mcitedefaultseppunct}\relax
\EndOfBibitem
\bibitem[Villegas and Tavares(2010)Villegas, and Tavares]{Villegas2010}
Villegas,~C. E.~P.; Tavares,~M. R.~S. \emph{Applied Physics Letters}
  \textbf{2010}, \emph{96}, 186101\relax
\mciteBstWouldAddEndPuncttrue
\mciteSetBstMidEndSepPunct{\mcitedefaultmidpunct}
{\mcitedefaultendpunct}{\mcitedefaultseppunct}\relax
\EndOfBibitem
\bibitem[Hartmann \latin{et~al.}(2010)Hartmann, Robinson, and
  Portnoi]{Hartmann2010}
Hartmann,~R.~R.; Robinson,~N.~J.; Portnoi,~M.~E. \emph{Physical Review B}
  \textbf{2010}, \emph{81}, 245431\relax
\mciteBstWouldAddEndPuncttrue
\mciteSetBstMidEndSepPunct{\mcitedefaultmidpunct}
{\mcitedefaultendpunct}{\mcitedefaultseppunct}\relax
\EndOfBibitem
\bibitem[Wu(2011)]{Wu2011}
Wu,~Z. \emph{Applied Physics Letters} \textbf{2011}, \emph{98}, 082117\relax
\mciteBstWouldAddEndPuncttrue
\mciteSetBstMidEndSepPunct{\mcitedefaultmidpunct}
{\mcitedefaultendpunct}{\mcitedefaultseppunct}\relax
\EndOfBibitem
\bibitem[Stone \latin{et~al.}(2012)Stone, Downing, and Portnoi]{Stone2012}
Stone,~D.~a.; Downing,~C.~a.; Portnoi,~M.~E. \emph{Physical Review B}
  \textbf{2012}, \emph{86}, 075464\relax
\mciteBstWouldAddEndPuncttrue
\mciteSetBstMidEndSepPunct{\mcitedefaultmidpunct}
{\mcitedefaultendpunct}{\mcitedefaultseppunct}\relax
\EndOfBibitem
\bibitem[Williams \latin{et~al.}(2011)Williams, Low, Lundstrom, and
  Marcus]{Williams2011}
Williams,~J.~R.; Low,~T.; Lundstrom,~M.~S.; Marcus,~C.~M. \emph{Nature
  Nanotechnology} \textbf{2011}, \emph{6}, 222--225\relax
\mciteBstWouldAddEndPuncttrue
\mciteSetBstMidEndSepPunct{\mcitedefaultmidpunct}
{\mcitedefaultendpunct}{\mcitedefaultseppunct}\relax
\EndOfBibitem
\bibitem[Wang \latin{et~al.}(2008)Wang, Ouyang, Li, Wang, Guo, and
  Dai]{Wang2008}
Wang,~X.; Ouyang,~Y.; Li,~X.; Wang,~H.; Guo,~J.; Dai,~H. \emph{Physical Review
  Letters} \textbf{2008}, \emph{100}, 206803\relax
\mciteBstWouldAddEndPuncttrue
\mciteSetBstMidEndSepPunct{\mcitedefaultmidpunct}
{\mcitedefaultendpunct}{\mcitedefaultseppunct}\relax
\EndOfBibitem
\bibitem[Tombros \latin{et~al.}(2011)Tombros, Veligura, Junesch, Guimar\~{a}es,
  Marun, Jonkman, and van Wees]{Tombros22011}
Tombros,~N.; Veligura,~A.; Junesch,~J.; Guimar\~{a}es,~M. H.~D.; Marun,~I.
  J.~V.; Jonkman,~H.~T.; van Wees,~B.~J. \emph{Nature Physics} \textbf{2011},
  \emph{7}, 697--700\relax
\mciteBstWouldAddEndPuncttrue
\mciteSetBstMidEndSepPunct{\mcitedefaultmidpunct}
{\mcitedefaultendpunct}{\mcitedefaultseppunct}\relax
\EndOfBibitem
\bibitem[Baringhaus \latin{et~al.}(2014)Baringhaus, Ruan, Edler, Tejeda, Sicot,
  Taleb-Ibrahimi, Li, Jiang, Conrad, Berger, Tegenkamp, and
  de~Heer]{Baringhaus2014}
Baringhaus,~J.; Ruan,~M.; Edler,~F.; Tejeda,~A.; Sicot,~M.; Taleb-Ibrahimi,~A.;
  Li,~A.-P.; Jiang,~Z.; Conrad,~E.~H.; Berger,~C.; Tegenkamp,~C.;
  de~Heer,~W.~A. \emph{Nature} \textbf{2014}, \emph{506}, 349--354\relax
\mciteBstWouldAddEndPuncttrue
\mciteSetBstMidEndSepPunct{\mcitedefaultmidpunct}
{\mcitedefaultendpunct}{\mcitedefaultseppunct}\relax
\EndOfBibitem
\bibitem[Goossens \latin{et~al.}(2012)Goossens, Driessen, Baart, Watanabe,
  Taniguchi, and Vandersypen]{Goossens2012}
Goossens,~A. S.~M.; Driessen,~S. C.~M.; Baart,~T.~A.; Watanabe,~K.;
  Taniguchi,~T.; Vandersypen,~L. M.~K. \emph{Nano Letters} \textbf{2012},
  \emph{12}, 4656--4660, PMID: 22906072\relax
\mciteBstWouldAddEndPuncttrue
\mciteSetBstMidEndSepPunct{\mcitedefaultmidpunct}
{\mcitedefaultendpunct}{\mcitedefaultseppunct}\relax
\EndOfBibitem
\bibitem[Allen \latin{et~al.}(2012)Allen, Martin, and Yacoby]{Allen2012}
Allen,~M.~T.; Martin,~J.; Yacoby,~A. \emph{Nature Comm} \textbf{2012},
  \emph{3}, 934\relax
\mciteBstWouldAddEndPuncttrue
\mciteSetBstMidEndSepPunct{\mcitedefaultmidpunct}
{\mcitedefaultendpunct}{\mcitedefaultseppunct}\relax
\EndOfBibitem
\bibitem[Williams and Marcus(2011)Williams, and Marcus]{Williams22011}
Williams,~J.~R.; Marcus,~C.~M. \emph{Physical Review Letters} \textbf{2011},
  \emph{107}, 046602\relax
\mciteBstWouldAddEndPuncttrue
\mciteSetBstMidEndSepPunct{\mcitedefaultmidpunct}
{\mcitedefaultendpunct}{\mcitedefaultseppunct}\relax
\EndOfBibitem
\bibitem[Rickhaus \latin{et~al.}(2015)Rickhaus, Makk, Liu, T\'ov\'ari, Weiss,
  Maurand, Richter, and Sch\"{o}nenberger]{Rickhaus2015}
Rickhaus,~P.; Makk,~P.; Liu,~M.-H.; T\'ov\'ari,~E.; Weiss,~M.; Maurand,~R.;
  Richter,~K.; Sch\"{o}nenberger,~C. \emph{Nature Communications}
  \textbf{2015}, \emph{6}, 6470\relax
\mciteBstWouldAddEndPuncttrue
\mciteSetBstMidEndSepPunct{\mcitedefaultmidpunct}
{\mcitedefaultendpunct}{\mcitedefaultseppunct}\relax
\EndOfBibitem
\bibitem[Taychatanapat \latin{et~al.}(2015)Taychatanapat, Tan, Yeo, Watanabe,
  Taniguchi, and \"{O}zyilmaz]{Taychatanapat2015}
Taychatanapat,~T.; Tan,~J.~Y.; Yeo,~Y.; Watanabe,~K.; Taniguchi,~T.;
  \"{O}zyilmaz,~B. \emph{Nature Communications} \textbf{2015}, \emph{6},
  6093\relax
\mciteBstWouldAddEndPuncttrue
\mciteSetBstMidEndSepPunct{\mcitedefaultmidpunct}
{\mcitedefaultendpunct}{\mcitedefaultseppunct}\relax
\EndOfBibitem
\bibitem[Young and Kim(2009)Young, and Kim]{Young2009}
Young,~A.~F.; Kim,~P. \emph{Nature Physics} \textbf{2009}, \emph{5},
  222--226\relax
\mciteBstWouldAddEndPuncttrue
\mciteSetBstMidEndSepPunct{\mcitedefaultmidpunct}
{\mcitedefaultendpunct}{\mcitedefaultseppunct}\relax
\EndOfBibitem
\bibitem[Cheianov and Falko(2006)Cheianov, and Falko]{Cheianov2006}
Cheianov,~V.~V.; Falko,~V.~I. \emph{Physical Review B} \textbf{2006},
  \emph{74}, 041403 (R)\relax
\mciteBstWouldAddEndPuncttrue
\mciteSetBstMidEndSepPunct{\mcitedefaultmidpunct}
{\mcitedefaultendpunct}{\mcitedefaultseppunct}\relax
\EndOfBibitem
\bibitem[Tombros \latin{et~al.}(2011)Tombros, Veligura, Junesch, van~den Berg,
  Zomer, Vera-Marun, Jonkman, and van Wees]{Tombros2011}
Tombros,~N.; Veligura,~A.; Junesch,~J.; van~den Berg,~J.; Zomer,~P.;
  Vera-Marun,~I.; Jonkman,~H.; van Wees,~B. \emph{Journal of Applied Physics}
  \textbf{2011}, \emph{109}, 093702\relax
\mciteBstWouldAddEndPuncttrue
\mciteSetBstMidEndSepPunct{\mcitedefaultmidpunct}
{\mcitedefaultendpunct}{\mcitedefaultseppunct}\relax
\EndOfBibitem
\bibitem[Maurand \latin{et~al.}(2014)Maurand, Rickhaus, Makk, Hess, Tovari,
  Handschin, Weiss, and Sch\"{o}nenberger]{Maurand2014}
Maurand,~R.; Rickhaus,~P.; Makk,~P.; Hess,~S.; Tovari,~E.; Handschin,~C.;
  Weiss,~M.; Sch\"{o}nenberger,~C. \emph{Carbon} \textbf{2014}, \emph{79},
  486\relax
\mciteBstWouldAddEndPuncttrue
\mciteSetBstMidEndSepPunct{\mcitedefaultmidpunct}
{\mcitedefaultendpunct}{\mcitedefaultseppunct}\relax
\EndOfBibitem
\bibitem[Liu \latin{et~al.}(2015)Liu, Rickhaus, Makk, T\'{o}v\'{a}ri, Maurand,
  Tkatschenko, Weiss, Sch\"{o}nenberger, and Richter]{Liu2015}
Liu,~M.-H.; Rickhaus,~P.; Makk,~P.; T\'{o}v\'{a}ri,~E.; Maurand,~R.;
  Tkatschenko,~F.; Weiss,~M.; Sch\"{o}nenberger,~C.; Richter,~K. \emph{Physical
  Review Letters} \textbf{2015}, \emph{114}, 036601\relax
\mciteBstWouldAddEndPuncttrue
\mciteSetBstMidEndSepPunct{\mcitedefaultmidpunct}
{\mcitedefaultendpunct}{\mcitedefaultseppunct}\relax
\EndOfBibitem
\bibitem[Freitag \latin{et~al.}(2012)Freitag, Weiss, Maurand, Trbovic, and
  Sch\"{o}nenberger]{Freitag2012}
Freitag,~F.; Weiss,~M.; Maurand,~R.; Trbovic,~J.; Sch\"{o}nenberger,~C.
  \emph{Solid State Communications} \textbf{2012}, \emph{152}, 2053--2057\relax
\mciteBstWouldAddEndPuncttrue
\mciteSetBstMidEndSepPunct{\mcitedefaultmidpunct}
{\mcitedefaultendpunct}{\mcitedefaultseppunct}\relax
\EndOfBibitem
\bibitem[Liu \latin{et~al.}(2012)Liu, Bundesmann, and Richter]{Liu22012}
Liu,~M.-H.; Bundesmann,~J.; Richter,~K. \emph{Physical Review B} \textbf{2012},
  \emph{85}, 085406\relax
\mciteBstWouldAddEndPuncttrue
\mciteSetBstMidEndSepPunct{\mcitedefaultmidpunct}
{\mcitedefaultendpunct}{\mcitedefaultseppunct}\relax
\EndOfBibitem
\end{mcitethebibliography}

\newpage
\section{Supporting Information}

\renewcommand{\figurename}{Supporting Figure}
\setcounter{figure}{0}    
\subsection{Measurements on further devices} 

\begin{figure}[htbp]
\includegraphics[width=1\columnwidth]{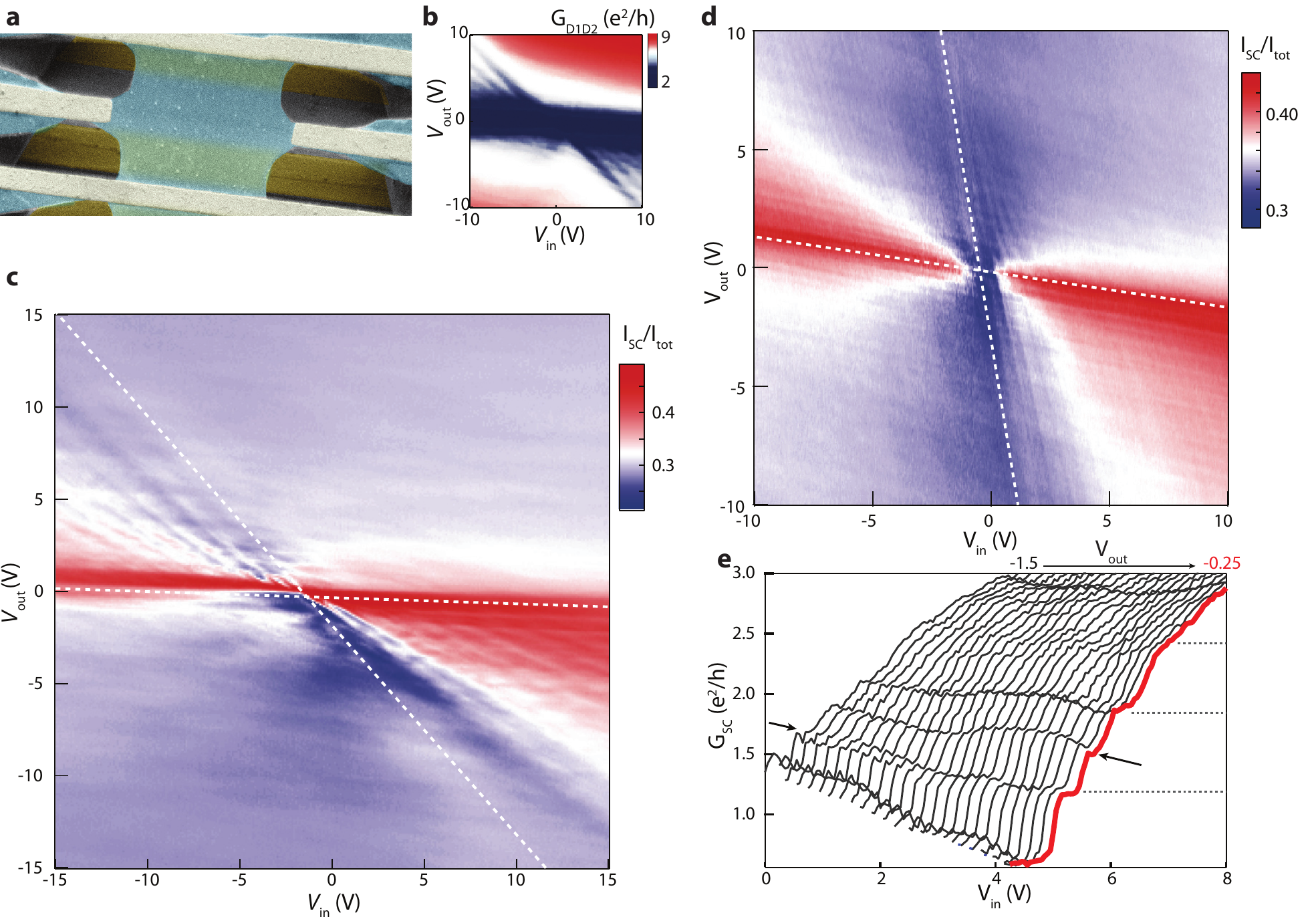}
\caption{\textbf{New measurement of electron guiding.}
\textbf{a}, Scanning electron micrograph of an etched guiding device.
\textbf{b} The Fabry-P\'erot map $G_{\rm D1D2}(V_{\rm in},V_{\rm out})$ shows a low conductance in the outer region of the device.
\textbf{c} Map of guiding efficiency $I_{SC}/I_{tot}$. The non-guiding value in the unipolar regime is around 27\%, in the optical-fibre-guiding regime (e.g. at $V_{\rm in}=15$V and $V_{\rm out}=2$V) the efficiency climbs to 32\% and in the pn regime ($V_{\rm in}=15$V and $V_{\rm out}=-2$V) it increases by 22\% to 49\%.
\textbf{d}, Guiding efficiency map of a third device with less crosstalk between the gates. The guiding efficiency is comparable to the former devices. OF guiding is clearly visible in this case.
\textbf{e}, Here, as in the device shown in the main text, steps as a function channel density appear in the conductance $G_{\rm SC}(V_{\rm in})$.
}
\label{guid:suppfig1}
\end{figure}

In an effort to reduce the injection losses into the channel we also fabricated samples that were etched at the sides, such as the one shown in Supporting Figure~\ref{guid:suppfig1}a. Here the injection into the channel is better defined, but unfortunately after current annealing, the device is not uniformly clean. In the Fabry-P\'erot map $G_{\rm D1D2}$ (Supporting Figure \ref{guid:suppfig1}b, the outer region exhibits a much stronger depletion of carrier densities (lower conductance) at the Dirac point, i.e. the lowest conductance is found at $V_{\rm out}=0$. 

Compared to the device shown in the main text, the one shown here exhibits a slightly higher absolute guiding efficiency of 49\% compared to 42\% before (Supporting Figure~\ref{guid:suppfig1}c). This is due to the optimized injection into the channel. The relative increase from the unipolar value of 27\% to 49\% in the p-n-guiding regime is with 81\% much higher than the 50\% relative increase in the older sample. The device was destructed by further attempts of current annealing. On the other hand, the OF guiding is much weaker than for the device shown in the main text. In addition, steps due to mode filling can not be clearly observed in this device. This is also due to a different geometry: In this case, the channel was formed by three bottomgate electrodes instead of one electrode and a global backgate as in the main text. The designed channel width was with 600 nm much larger than in the main text, which further complicates the observation of single mode-filling. We conclude that, even though the device was in the channel region not as clean as the device shown in the main text, the injection properties were improved due to the optimized design.

The measured guiding efficiency of a third device is shown in Supporting Figure \ref{guid:suppfig1}d. For geometrical reasons, this device exhibits less crosstalk between the gates, i.e. the white dashed lines are less tilted in this case. Even though the overall guiding efficiency is slightly lower, OF guiding and steps in the conductance $G_{\rm SC}(V_{\rm in})$ are visible. The features appear to be similar to the mode filling steps described in the main text, i.e. almost equally spaced plateaus and additional resonances. With dashed lines we indicate features that are probably due to mode filling, and with arrows we mark what might be the two-cavity resonances. The plateaus are however less constant with increasing $V_{\rm out}$. In order to distinguish mode filling from coherent two-cavity resonances, detailed comparison to a modified theoretical model would be required.

\subsection{Voltage to density mapping}
\begin{figure}[htbp]
\includegraphics[width=1\columnwidth]{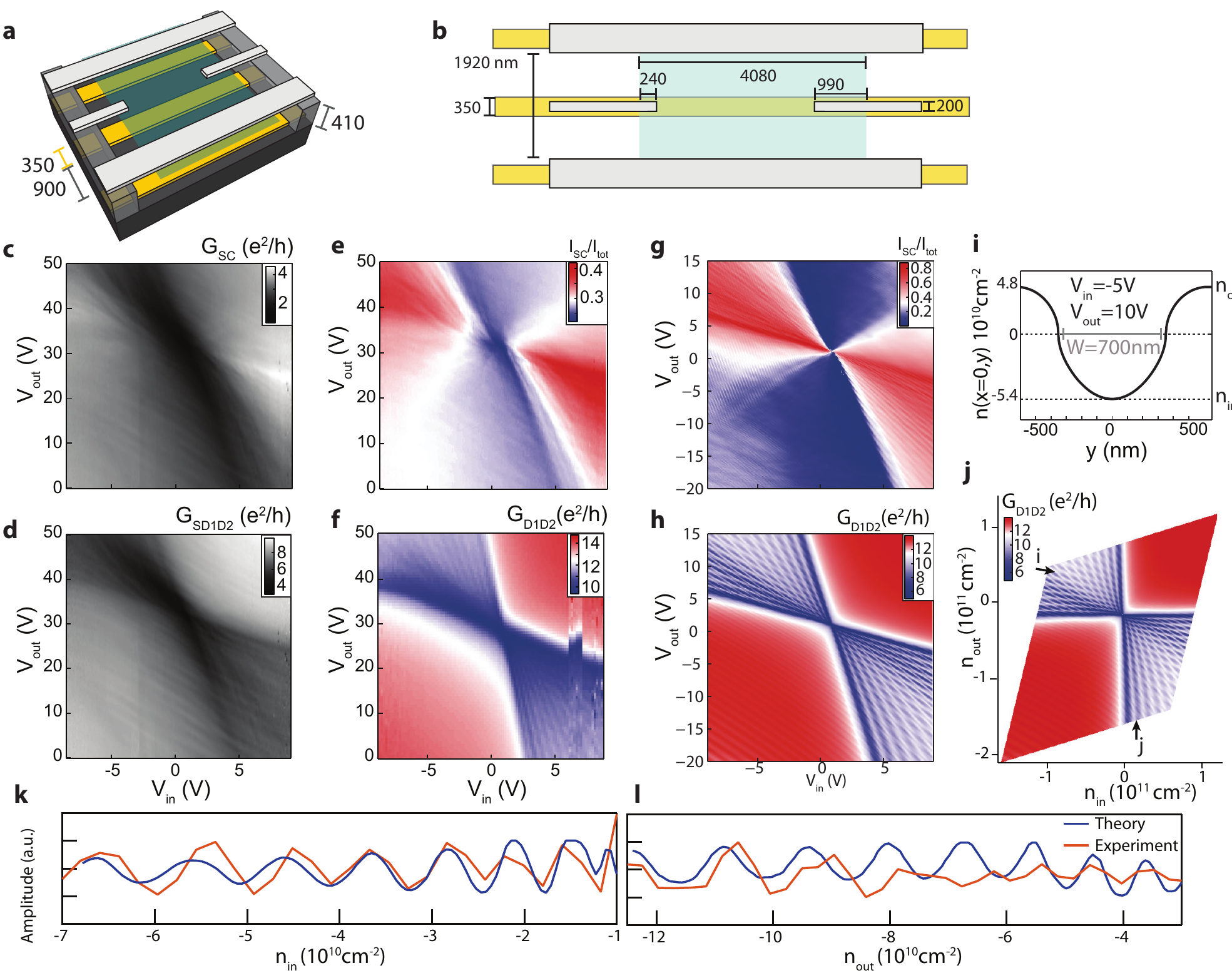}
\caption{\textbf{Original data and density mapping.}
\textbf{a-b}, 3D and top-view of the measured device with its dimensions used also for the simulations.
\textbf{c}, The measured guiding conductance $G_{\rm SC}$ between left and right contact,
\textbf{d}, and the simultaneously measured loss $G_{\rm SD1D2}$ between S and D1 and D2 contact.
\textbf{e}, The guiding efficiency $\gamma$ is calculated from these two maps as described in the main text, i.e. $\gamma=G_{\rm SC}/(G_{\rm SC}+G_{\rm SD1D2})$.
\textbf{f}, The Fabry-P\'erot map is measured in a different configuration. 
\textbf{g}, The simulated guiding efficiency and the simulated Fabry-P\'erot pattern \textbf{h} show the same crosstalk between the inner and outer gate. The voltage scale for the outer voltage $V_{\rm out}$ differs from the experimental data, but not the $V_{\rm in}$ scale. This is due to a problem with the backgate, i.e. the experimental $V_{\rm out}$ scale is not trustworthy.
\textbf{i}, The density mapping is done according to calculated density profiles as shown in this Figure.
\textbf{j}, The $G_{\rm D1D2}(V_{\rm in},V_{\rm out})$ to $G_{\rm D1D2}(n_{\rm in}, n_{\rm out})$ mapping shears the data since the capacitive crosstalk is corrected.
 \textbf{k,l}, The comparison of the Fabry-P\'erot oscillation period between experiment and theory reveals a successful voltage to density mapping. Plotted are the oscillations as a function of $n_{\rm in}$ in \textbf{k} and as a function of $n_{\rm out}$ in \textbf{l} along the directions marked with arrows in \textbf{j}. The curves were normalized after subtraction of the background conductance.
}
\label{guid:suppfig2} 
\end{figure}

The detailed device geometry as used for the simulations is shown in Supporting Figure\ref{guid:suppfig2}a and b.
The measurement setup for the guiding efficiency is explained in the main text: An AC voltage is applied to the source (S) contact and an AC current is simultaneously measured at the collector (C) and the side drain contacts (D1 and D2). In Supporting Figure \ref{guid:suppfig2}c we show the measured conductance between the S and C contact $G_{\rm SC}=I_{\rm C}/V_{\rm S}$ and in Supporting Figure \ref{guid:suppfig2}d the conductance $G_{\rm SD1D2}=(I_{\rm D1}+I_{\rm D2})/V_{\rm S}$ between the source and side drain contacts. The guiding efficiency in Supporting Figure \ref{guid:suppfig2}e is then given by $\gamma=I_{\rm SC}/(I_{\rm SD1}+I_{\rm SD2}+I_{\rm SC})=G_{\rm SC}/(G_{\rm SC}+G_{\rm SD1D2})$. 

Apparently the $V_{\rm out}$ scale for $\gamma$ and for the Fabry-P\'erot map $G_{\rm D1D2}$ (Supporting Figure \ref{guid:suppfig2}f) is shifted to high voltages. Whereas the global Dirac point is very close to $V_{\rm in}=0$V, it occurs at roughly $V_{\rm out}=30$V for the outer gate. The small shift in $V_{\rm in}$ and the regular Fabry-P\'erot pattern clearly demonstrate that the shift in $V_{\rm out}$ is not due to doping of the graphene but rather due to experimental problems with the backgate, that also lead to gate-jumps in the Fabry-P\'erot map (Supporting Figure \ref{guid:suppfig2}f). The backgate was connected with a silver-paint and this connection might have become bad at low temperatures, leading to charging of the gates and to instabilities. We therefore discard the $V_{\rm out}$-scale since the values are apparently not trustworthy, while keeping the $V_{\rm in}$-scale. 

By comparing to the simulated guiding efficiency (Supporting Figure \ref{guid:suppfig2}g) and Fabry-P\'erot map (Supporting Figure \ref{guid:suppfig2}h) the correct density values for the experimental data can be extracted. This is possible since there are no free fitting parameters in the model and all geometrical parameters which do influence the capacitive crosstalk between the gates are fixed. In addition, the $V_{\rm in}$ voltage scale can be compared. Both experimental and theoretical data are mapped from $(V_{\rm in},V_{\rm out})$ to $(n_{\rm in}, n_{\rm out})$ by using the local density profiles $n(x=0,y)$ (Supporting Figure \ref{guid:suppfig2}i). The maximal absolute density in the channel $n_{\rm in}$ and of the maximal absolute density in the outer region $n_{\rm out}$ are obtained for all voltages $(V_{\rm in},V_{\rm out})$. Such a mapping is done for the theoretical Fabry-P\'erot map in Supporting Figure \ref{guid:suppfig2}j which is not shown in the main text. The data are sheared since the capacitive crosstalk between backgate and bottomgate is corrected.

The spacing of the Fabry-P\'erot resonances represents a good control for the voltage to density mapping. Since the potential is smooth and the cavity size is changing with applied gate voltage, the experimental Fabry-P\'erot resonances are best compared to the oscillations obtained in the tight-binding simulation, as it is done in Supporting Figure \ref{guid:suppfig2}k for the density in the channel and in Supporting Figure \ref{guid:suppfig2}l for $n_{\rm out}$. As the periodicity of the measured and calculated oscillation is matching rather well, we conclude that the density is properly mapped. The density of two resonance maxima $n_i$ and $n_{i+1}$ are spaced $\sqrt{n_{i+1}}-\sqrt{n_{i}}=\sqrt{\pi}/W$, where $W$ is the width of the channel. This can be obtained using the interference condition $k_FW=N\pi$ with $N$ an integer.
From the two first maxima in Supporting Figure \ref{guid:suppfig2}k (experimental or theoretical) a cavity size of $730\unit{nm}$ is obtained, which corresponds well to the cavity size obtained from electrostatic simulations (see Supporting Figure \ref{guid:suppfig2}i).

\subsection{Smooth versus sharp p-n interface}
\begin{figure}[htbp]
\includegraphics[width=0.8\textwidth]{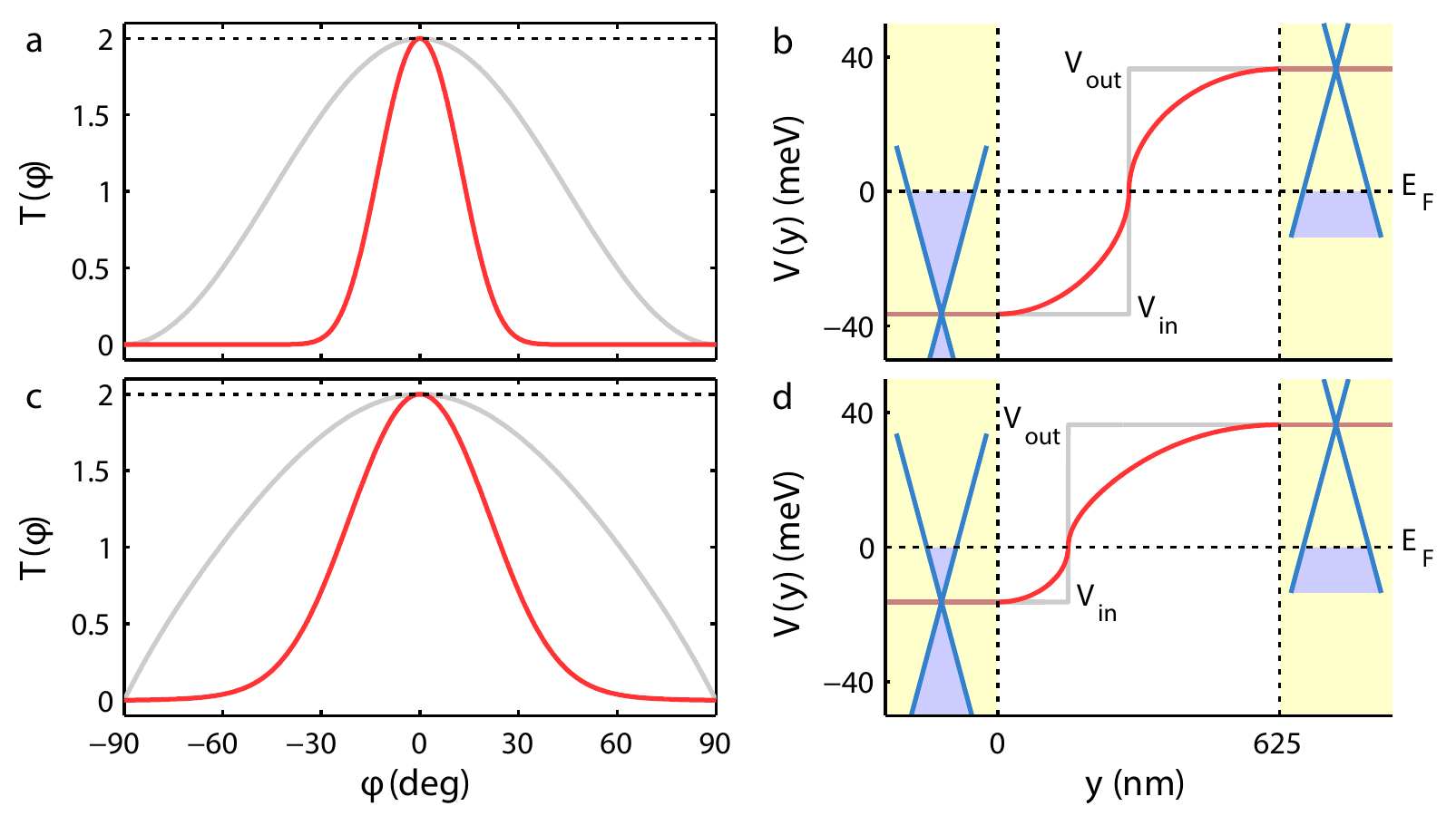}
\caption{Angle-resolved transmission function (red curves) $T(\phi)$ across the pn junction between the guiding channel center and the outer cavity at carrier densities $(n_{\rm in},n_{\rm out})=(10,-10)\times 10^{10}\unit{cm}^{-2}$ in \textbf{a} and $(n_{\rm in},n_{\rm out})=(2,-10)\times 10^{10}\unit{cm}^{-2}$ in \textbf{c}, taking into account the smooth junction profiles $V(y)$ obtained from the electrostatics and sketched (red curves) in \textbf{b} and \textbf{d}, respectively. Gray curves in \textbf{a} and \textbf{c} are $T(\phi)$ when the smooth $V(y)$'s are replaced with ideally abrupt profiles as sketched (gray steps) in \textbf{b} and \textbf{d}, where $V_{\rm in/out}=-\func{sgn}(n_{\rm in/out})\hbar v_F\sqrt{\pi|n_{\rm in/out}|}$ is the band offset applied on the incoming/outgoing lead (shaded in light yellow).}
\label{guid:suppfig3}
\end{figure}

To highlight the contrast between the smooth p-n junction of our suspended graphene and the widely discussed abrupt junction, we show in Supporting Figure \ref{guid:suppfig3} the angle-resolved transmission $T(\phi)$ across a p-n junction, considering the realistic smooth potential $V(y)$ obtained from the electrostatic simulation, as well as the abrupt case for comparison. The red curves correspond to a smooth potential step, the gray one to a sharp p-n interface. The $T(\phi)$ curves shown in Supporting Figure \ref{guid:suppfig3}a and \ref{guid:suppfig3}c consider symmetric and asymmetric pn junctions, respectively, and are calculated in the same way as ref.~\citenum{Liu22012}. In the symmetric case of Supporting Figure \ref{guid:suppfig3}a,b with $n_{\rm in}=-n_{\rm out}=10^{11}\unit{cm}^{-2}$, the transmission curve for the abrupt case corresponds to the well-known expression\cite{Cheianov2006}, $T(\phi)=\cos^2(\phi)$, so that $T$ drops to one half at $\pm45^\circ$, while for the smooth case of our device $T$ drops to one half (which is $T=1$ here because the valley degeneracy is automatically included in the tight-binding calculation\cite{Liu22012}) at about $\pm14^\circ$. In the asymmetric case of Supporting Figure \ref{guid:suppfig3}c,d with $|n_{\rm in}|<|n_{\rm out}|$, the peak of $T(\phi)$ for the smooth case remains relatively narrow, so that guided modes at such a $(n_{\rm in},n_{\rm out})$ configuration outside the OF guiding regime become possible.

\subsection{Two-cavity resonances}
\begin{figure}
\includegraphics[width=1\textwidth]{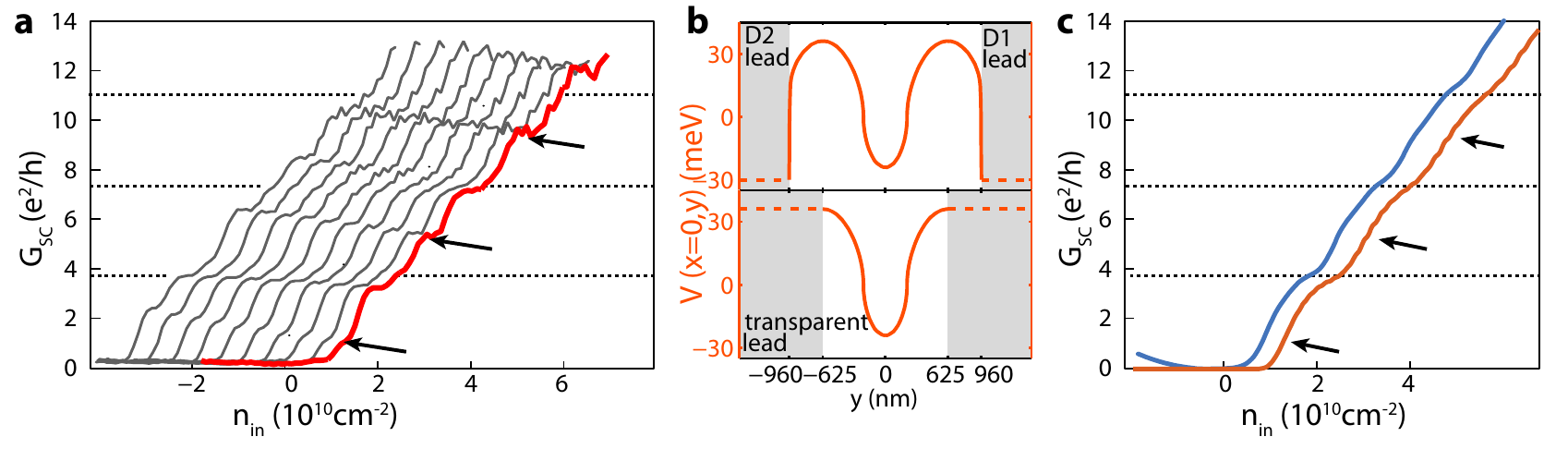}
\caption{
\textbf{The two-cavity resonances disappear once the outer leads D1 and D2 are transparent.}
\textbf{a}, A similar plot as shown in the main text (Figure 5c) showing calculated conductance $G_{\rm SC}$. The plateaus due to mode-filling occur along the dashed grid lines, the two-cavity resonances are marked with black arrows.
\textbf{b}, Potential profiles $V(x=0,y)$ as used in the main text (upper) and idealized such that reflections at the outer D1 and D2 contacts are suppressed (lower). 
\textbf{c}, For an idealized profile the two-cavity resonances do not occur anymore. The plateaus for mode-filling are seen for a 3 $\mu$m (orange) and a 500 nm (blue) long channel.}
\label{guid:suppfig4}
\end{figure}

In the main text we mention that the two-cavity resonances, which we mark with black arrows in Figures 4a, 4
b and Supporting Figure \ref{guid:suppfig4}a are due to resonances that are tuned by both densities $n_{\rm in}$ and $n_{\rm out}$. For these resonances, reflections at the side contacts D1 and D2 are required. By modifying the potential profile as shown in Supporting Figure \ref{guid:suppfig4}b and simulating transparent (non-reflective) leads, the two-cavity resonances on the $G_{\rm SC}$ curves indeed disappear (Supporting Figure \ref{guid:suppfig4}c). The position where they appear in the realistic device simulation are marked with black arrows. The shoulders close to 4,8 and 12$e^2/h$ remain clearly visible.
In this idealized simulation, the source and collector contacts are set to be 3$\mu$m apart similar to the real device (orange curve). Additionally we also show a test where the S and C contacts are 500 nm apart (blue curve).

\subsection{Bandstructures}
\begin{figure}[h]
\includegraphics[width=1\textwidth]{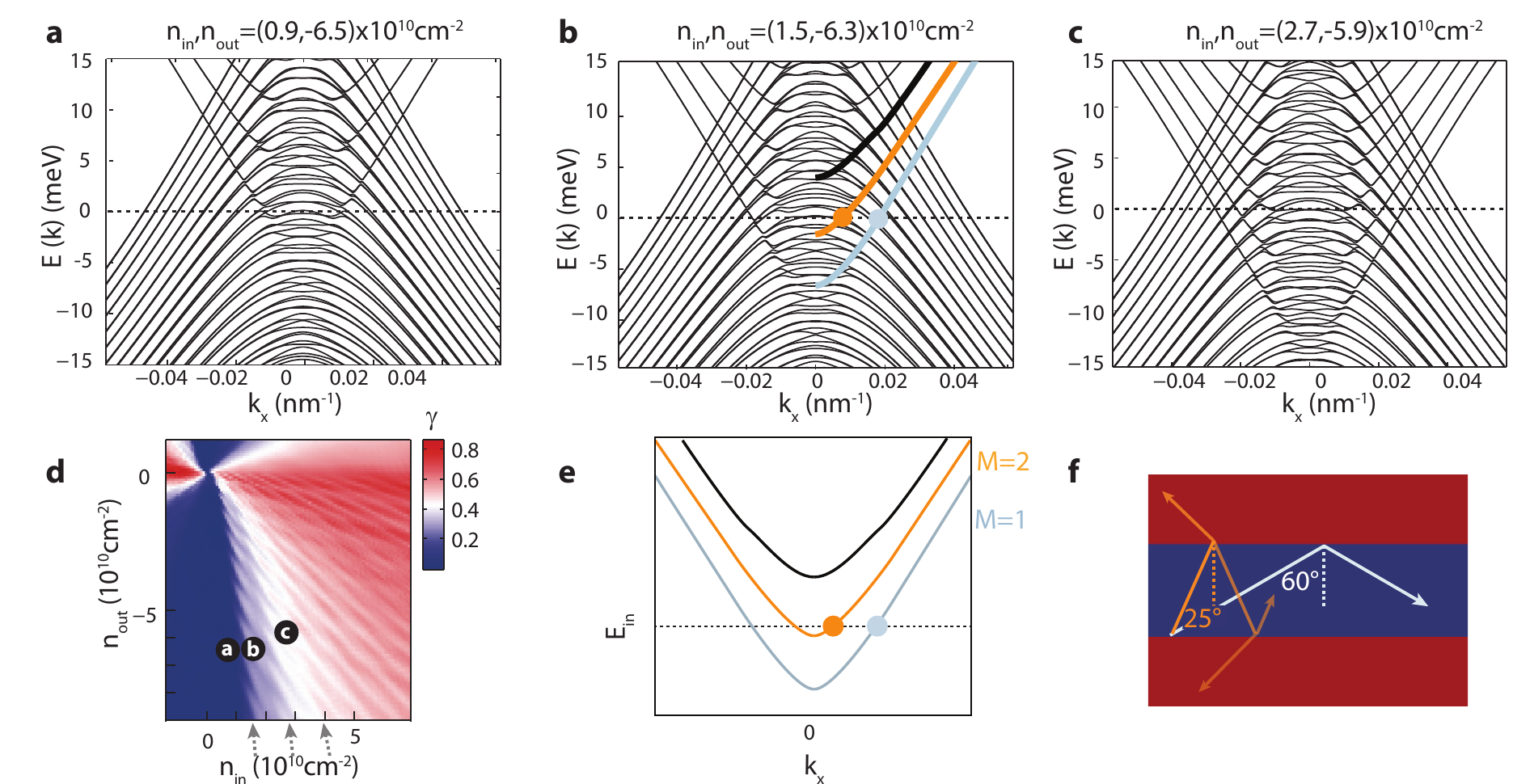}
\caption{
\textbf{Calculated band structures for translational invariance along x.}
\textbf{a-c}, Band structures $E(k_{\rm x})$ in the empty-channel regime (a), when the first mode appears in transport (b) and for two modes (c).
\textbf{d}, The corresponding positions are labeled on the calculated $\gamma(n_{\rm in},n_{\rm out})$ map.
\textbf{e}, A simplified picture of b, also shown in Figure 1f of the main text.
\textbf{f}, The maximum angles of incidence to the p-n interface are sketched for the two different modes.
}
\label{guid:suppfig5}
\end{figure}

In Figure 1f in the main text we sketched a very much simplified band structure for the modes in order to illustrate the leakage of the guide modes with short $k_x$. A more evolved picture is presented in Supporting Figure \ref{guid:suppfig5}a-c, where we show band structures $E(k_x)$ calculated by taking a unit cell laterally cut (along $y$) from the simulated graphene lattice at around $x=0$. The flake is oriented such that the armchair edge is parallel to the p-n interface. The band structures are given for different densities $n_{\rm in},n_{\rm out}$ and the corresponding position is labeled in the guiding efficiency map, shown in Supporting Figure \ref{guid:suppfig5}d. We observe zero guided mode in transport along the channel in Supporting Figure \ref{guid:suppfig5}a, one mode in b, and two in c. The band structures consist of mainly two Dirac-like-cones. The one for hole-like transport at the Fermi energy corresponds to the outer cavities. Since $n_{\rm out}$ is changing only little from Supporting Figure \ref{guid:suppfig5}a-c, this part of the band-structure remains unchanged. Due to the rather high density, there is a large number of modes available. More interesting are the modes that appear in the inner cavity, corresponding to electron-like transport. In Supporting Figure \ref{guid:suppfig5}b we overlay curves that lead to the simplified picture shown in Figure 1f and Supporting Figure \ref{guid:suppfig5}e by neglecting the fine-structure at low $k_x$. We would expect to observe two modes in transport along the channel for Supporting Figure \ref{guid:suppfig5}b due to the orange and the blue branch. However, by calculating $k_y=\sqrt{k^2-k_x^2}$ for $k=\sqrt{n_{\rm in}\pi}=0.022$ nm$^{-1}$ we obtain an angle of incidence of 60$^{\circ}$ for the blue and only 25$^{\circ}$ for the orange mode. Comparing to the numbers given in Supporting Figure \ref{guid:suppfig3}a and c for transmission at a smooth p-n interface it becomes apparent why in transport along the channel only one of the modes is seen: the smaller angle of incidence for the second (orange) mode leads to leakage out of the channel, whereas the first (blue) mode can be guided efficiently.

%
%


\end{document}